%% file: samplepaper.tex
\documentclass[runningheads]{llncs}
\usepackage[T1]{fontenc}

\usepackage{marvosym}
\newcommand{\corrauth}{\mbox{\normalfont\Letter}}
\usepackage{graphicx}
\usepackage{booktabs}   
\usepackage{graphicx}   
\usepackage{multirow}   
\usepackage{amsmath}   
\usepackage{amsfonts}  
\usepackage{xcolor}
\usepackage{colortbl}
\usepackage{marvosym} 
\newcommand{\PM}[2]{#1{\scriptsize$\pm$#2}}

\newcommand{\BEST}[1]{\textcolor{red}{\underline{#1}}}

\newcommand{\SND}[1]{\textcolor{blue}{#1}}

\newcommand{\GAIN}[1]{\textcolor{green!45!black}{#1}}
\newcommand{\DeltaMark}{\textcolor{green!45!black}
{$\Delta$}}

\newcommand{\greencheck}{\textcolor{green!70!black}{\checkmark}}

\begin{document}
\title{C$^2$LUE: Clinically Aligned Patient-Level Ultrasound Diagnosis via Evidence-Aware Multi-Instance Learning}
\titlerunning{C$^2$LUE}
\author{Qin-Yi Zhang\inst{1,2} \and
Linying Cao\inst{1,2} \and
De-Xing Huang\inst{1,2} \and
Haichuan Lin\inst{1,2} \and
Yi~Yang\inst{1,2} \and
Manzhi Wang\inst{1,2} \and
Xiao-Hu Zhou\inst{1,2} \and
Chen Chen\inst{1,2} \and
Shuangyi~Wang\inst{1,2(\corrauth)} \and
Zeng-Guang Hou\inst{1,2(\corrauth)}}
\authorrunning{Qin-Yi Zhang et al.}
\institute{State Key Laboratory of Multimodal Artificial Intelligence Systems,
Institute of Automation, Chinese Academy of Sciences, Beijing 100190, China \and
School of Artificial Intelligence, University of Chinese Academy of Sciences,
Beijing 100049, China\\
\email{\{shuangyi.wang, zengguang.hou\}@ia.ac.cn}}
\maketitle

\begin{abstract}
Patient-level ultrasound diagnosis demands not only high predictive accuracy but also clinically aligned mitigation of asymmetric diagnostic risks. While Multi-Instance Learning (MIL) effectively models routine multi-view exams, standard aggregators often dilute heterogeneous diagnostic cues and optimize solely for global accuracy. This frequently yields undesirable sensitivity--specificity trade-offs. To bridge this gap, we propose C$^2$LUE, a novel MIL framework coupling evidence-aware aggregation with risk-aligned regularization. To preserve evidence heterogeneity and prevent unimodal attention collapse, C$^2$LUE employs a multi-query cross-attention mechanism built upon a LoRA-adapted DINOv3 backbone. Multiple latent queries explicitly extract and disentangle diverse morphological patterns from the ultrasound sequence. Furthermore, to improve clinical risk alignment, we introduce a dynamic trade-off regularization objective that penalizes sensitivity--specificity imbalance during training. Rather than leaving the clinical trade-off entirely to post-hoc threshold adjustment, this objective encourages the learned score distribution toward a more balanced operating region. Extensive evaluations on patient-level, biopsy-proven thyroid and breast ultrasound datasets demonstrate that C$^2$LUE consistently outperforms prior alternatives. By uniting strong discrimination with a balanced error profile, C$^2$LUE provides a promising solution for real-world ultrasound screening and diagnosis.
\end{abstract}

\section{Introduction}
\label{sec:intro}

Ultrasound is a cornerstone of clinical screening and diagnosis due to its safety, cost-effectiveness, and real-time availability~\cite{van2019deep}. In routine workflows, physicians acquire multiple views or short sequences to formulate a single patient-level decision~\cite{qian2021prospective,AhmadiAmiri2024TMI_AS}. However, mapping these image sequences to an automated diagnosis presents a significant learning challenge. Diagnostic evidence is highly heterogeneous, encompassing diverse pathological signs such as irregular margins, subtle echogenicity changes, and microcalcifications~\cite{wan2021hierarchical}. Furthermore, these critical cues are sparsely distributed. The majority of the acquired frames are often redundant or heavily corrupted by operator variations, inconsistent probe angles, and speckle noise~\cite{freitas2024automatic}. Consequently, an effective diagnostic model must not only aggregate diverse evidence across instances, but also produce clinically useful risk scores that support a balanced sensitivity--specificity trade-off~\cite{tao2025enhancing}.

Multi-Instance Learning (MIL) offers a natural formulation for this setting by treating a patient's exam as a ``bag'' of instances~\cite{maron1997framework}. Yet, current MIL aggregators struggle with the morphological complexity of ultrasound. Global operators, such as mean or max pooling, inherently compromise performance by either diluting rare focal cues or becoming overly driven by a single high-scoring outlier~\cite{carbonneau2018multiple}. While attention-based pooling mitigates this issue, it frequently suffers from unimodal collapse—focusing exclusively on the most prominent lesion feature while suppressing complementary evidence~\cite{ilse2018attention,lee2019set,shao2021transmil}. In ultrasound, where distinct diagnostic signs are distributed across different anatomical views, this loss of evidence heterogeneity severely limits the model's diagnostic robustness~\cite{duong2025thyroidxl,liang2025medical}. 

Furthermore, achieving high average accuracy does not inherently translate to clinical utility~\cite{Wang2020MedIA_Thyroid,Zhao2022TMI_LGFD}. Cross-entropy optimization pushes the decision boundary toward the global minimum error, which often neglects the critical balance between false negatives (FN) and false positives (FP)~\cite{ren2020balanced,yuan2021large}. In screening and diagnostic scenarios, these errors carry highly asymmetric clinical costs: missing a malignancy risks delayed intervention, whereas excessive false alarms lead to unnecessary biopsies and patient anxiety~\cite{nelson2016harms}. While post-hoc threshold tuning is a common way to select clinical operating points, it can be brittle when the learned score distribution is skewed by class imbalance. Therefore, there is a pressing need for training-time regularization that better aligns model scores with clinical trade-off preferences~\cite{menon2021longtail,Huang2022MICCAI_PersonalizedThyroid}.

To bridge these gaps, we propose C$^2$LUE, a patient-level framework designed to jointly optimize evidence heterogeneity and clinical trade-offs. First, we address the aggregation challenge with a multi-query cross-attention module. Instead of a single attention bottleneck, multiple learnable queries act as distinct evidence extractors. This design explicitly preserves diverse diagnostic cues prior to classification. We support this aggregator using a parameter-efficient foundation model to capture robust representations from noisy ultrasound data. Second, we address objective misalignment by introducing a clinically aligned trade-off regularization. Rather than relying solely on post-hoc threshold adjustment, this objective dynamically penalizes sensitivity--specificity imbalance during training. Consequently, it encourages the learned scores toward a more balanced operating region while remaining adjustable to different clinical preferences.

Our main contributions are summarized as follows:
\begin{enumerate}
\renewcommand{\labelenumi}{(\arabic{enumi})}
    \item We propose C$^2$LUE, an evidence-aware MIL framework for patient-level ultrasound diagnosis. By replacing the single aggregation bottleneck with multi-query cross-attention, C$^2$LUE preserves heterogeneous diagnostic cues across ultrasound views and alleviates unimodal attention collapse.
    \item We introduce a clinically aligned regularization objective that incorporates the sensitivity--specificity trade-off into training. By penalizing metric imbalance at the patient level, this objective aligns the error profile with clinically balanced diagnosis.
    \item Evaluations on two biopsy-proven ultrasound benchmarks demonstrate that C$^2$LUE achieves the best mean patient-level performance among compared baselines. By combining strong discrimination with balanced sensitivity and specificity, it supports a more reliable operating point for patient-level ultrasound diagnosis.
\end{enumerate}

\section{Related Work}
\subsection{Patient-Level Ultrasound Diagnosis}

Deep learning has been widely explored for ultrasound diagnosis, including dynamic ultrasound interpretation, thyroid and breast lesion classification, and multi-view clinical decision support~\cite{wan2021hierarchical,Wang2020MedIA_Thyroid,Zhao2022TMI_LGFD,Yan2024CommMed_BreastUS}. However, many existing models still operate at the image or fixed-view level, whereas routine ultrasound diagnosis often requires physicians to integrate variable numbers of views or sequences before making a patient-level decision~\cite{qian2021prospective,Huang2022MICCAI_PersonalizedThyroid,Yan2024CommMed_BreastUS}. Multi-Instance Learning (MIL) naturally fits this setting by modeling each examination as a bag of instances~\cite{maron1997framework,carbonneau2018multiple,Lu2021NBE_CLAM}. Conventional pooling strategies may dilute sparse malignant cues or over-emphasize noisy outliers, while attention-based MIL and Transformer-based set models improve adaptive aggregation~\cite{ilse2018attention,lee2019set,shao2021transmil,li2021dual}. Nevertheless, most aggregators still compress heterogeneous evidence through a single representation bottleneck. For ultrasound, where diagnostic signs such as margins, echogenicity, and microcalcifications may appear across different views, such aggregation can suppress complementary cues and lead to insufficient evidence diversity~\cite{duong2025thyroidxl,liang2025medical}.

\subsection{Clinical Trade-off Alignment in Medical Diagnosis}

Beyond feature aggregation, clinical deployment requires a reliable operating point rather than merely high average accuracy~\cite{Guo2017ICML_Calibration,Rajaraman2022PLOSONE_Calibration,Geysels2025MICCAI_CalibrationMIL}. Standard cross-entropy optimization mainly minimizes global classification error, which may bias the decision boundary toward the majority class in imbalanced medical cohorts~\cite{ren2020balanced,menon2021longtail}. This is particularly problematic in cancer screening, where false negatives may delay intervention, while excessive false positives can lead to unnecessary biopsies and anxiety~\cite{nelson2016harms}. Existing remedies, such as class re-weighting, logit adjustment, imbalance-aware losses, and AUC-oriented objectives, alleviate distributional imbalance but do not directly enforce a balanced sensitivity--specificity trade-off~\cite{yuan2021large,menon2021longtail,Scholz2024MIDL_ImbalanceLoss}. Consequently, existing ultrasound diagnosis frameworks still lack an end-to-end mechanism that jointly preserves heterogeneous evidence and aligns patient-level predictions with clinically preferred sensitivity--specificity trade-offs during training.

\section{Methodology}
\label{sec:method}

\begin{figure*}[t]
    \centering
    \includegraphics[width=1.0\linewidth]{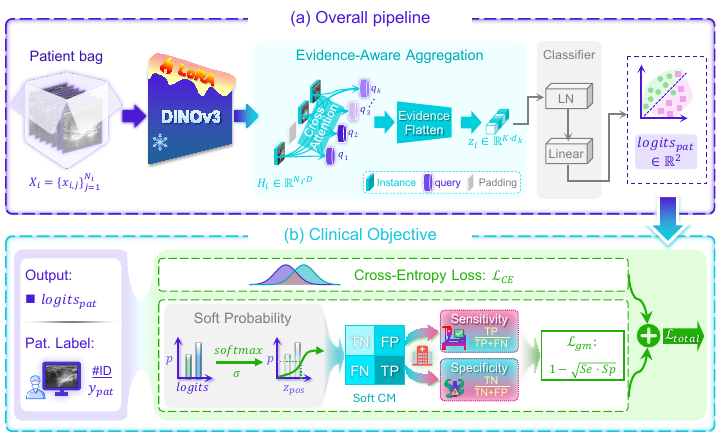}
    \caption{Overview of C$^2$LUE. (a) The evidence-aware pipeline extracts heterogeneous features via multi-query cross-attention on a LoRA-adapted DINOv3 backbone. (b) The clinical objective dynamically penalizes sensitivity-specificity imbalance to align the decision boundary with clinical trade-off preferences.}
    \label{fig:fig1}
\end{figure*}

Fig.~\ref{fig:fig1} illustrates the proposed C$^2$LUE architecture, which comprises two core pillars: (a) an evidence-aware learning pipeline for extracting heterogeneous diagnostic cues, and (b) a clinically aligned optimization objective for dynamically balancing the sensitivity--specificity trade-off.
\subsection{Evidence-Aware Learning Pipeline}
\label{subsec:pipeline}

\noindent\textbf{Feature Encoding via Low-Rank Adaptation.}
Given a patient ultrasound exam $X_i = \{x_{i,j}\}_{j=1}^{N_i}$ with $N_i$ instances, our objective is to predict the patient-level malignancy probability $P(Y_i=1|X_i)$. We employ a DINOv3 foundation model~\cite{simeoni2025dinov3}, fine-tuned via Low-Rank Adaptation (LoRA)~\cite{hu2022lora}, as the instance-level feature encoder. This projects the exam into a feature bag $H_i \in \mathbb{R}^{N_i \times D}$, where $D$ denotes the embedding dimension.

\noindent\textbf{Evidence Aggregation via Multi-Query Cross-Attention.}
Current MIL aggregators typically compress $H_i$ through a single attention bottleneck. This can lead to unimodal collapse, where the network focuses on the most salient feature while discarding complementary diagnostic evidence.

To explicitly preserve evidence heterogeneity, C$^2$LUE introduces a multi-query cross-attention aggregator. We instantiate $K$ learnable latent queries, denoted as $Q_{\mathrm{lat}} \in \mathbb{R}^{K \times D}$, which act as distinct evidence extractors and attend to the instance-level feature bag via cross-attention:
\begin{equation}
Z_i = \mathrm{MHA}(Q_{\mathrm{lat}}, H_i, H_i)
\in \mathbb{R}^{K \times D},
\end{equation}
where $\mathrm{MHA}(\cdot)$ denotes multi-head cross-attention with $Q_{\mathrm{lat}}$ as the query and $H_i$ as both key and value. In each attention head, scaled dot-product attention is computed in the projected head dimension, and the outputs of all heads are concatenated and projected back to the embedding dimension $D$. Therefore, the output $Z_i=\{z_{i,k}\}_{k=1}^{K}$ contains $K$ query-specific evidence vectors, where each $z_{i,k}\in\mathbb{R}^{D}$ is encouraged to capture complementary diagnostic patterns from the patient bag.

Because global average pooling across these $K$ vectors would mix query-specific evidence and weaken evidence diversity, we explicitly concatenate the query outputs to form a unified patient-level representation:
\begin{equation}
v_i = z_{i,1} \oplus z_{i,2} \oplus \dots \oplus z_{i,K}
\in \mathbb{R}^{K D},
\end{equation}
where $\oplus$ denotes vector concatenation. This flattening preserves the query-specific structure of the extracted evidence, allowing the classification head to independently weight distinct diagnostic patterns before computing the final patient-level logits.

\subsection{Clinically-Aligned Trade-off Regularization}
\label{subsec:objective}

Cross-Entropy (CE) optimization minimizes patient-level classification error, but in imbalanced medical cohorts it may favor the majority class and lead to an undesired sensitivity--specificity trade-off. To guide the model toward a more clinically balanced operating region, we introduce a differentiable Log-Geometric Mean regularization.

Let $\mathbf{y}, \mathbf{p} \in [0, 1]^B$ denote the ground-truth labels and predicted probabilities over a batch of $B$ patients, and $\mathbf{1}$ be a vector of ones. We first relax the discrete confusion matrix into continuous expectations via inner products:
\begin{align}
    TP &= \mathbf{y}^\top \mathbf{p}, \quad &FN &= \mathbf{y}^\top (\mathbf{1} - \mathbf{p}) \\
    TN &= (\mathbf{1} - \mathbf{y})^\top (\mathbf{1} - \mathbf{p}), \quad &FP &= (\mathbf{1} - \mathbf{y})^\top \mathbf{p}
\end{align}
The differentiable Sensitivity ($Se$) and Specificity ($Sp$) are then formulated as:
\begin{equation}
    Se = \frac{TP}{TP + FN + \epsilon}, \quad Sp = \frac{TN}{TN + FP + \epsilon}
\end{equation}
where $\epsilon$ acts as a numerical stabilizer.

To examine the class-wise weighting of the geometric-mean objective, we consider its logarithmic form. Defining $\mathcal{J} = \ln(Se+\epsilon) + \ln(Sp+\epsilon)$, the chain rule with respect to the network parameters $\theta$ yields:
\begin{equation}
    \nabla_\theta \mathcal{J}
    =
    \frac{1}{Se+\epsilon} \nabla_\theta Se
    +
    \frac{1}{Sp+\epsilon} \nabla_\theta Sp
\end{equation}
To analyze the local optimization dynamics, consider the gradient of $Se$ with respect to a single positive sample prediction $p_i$ where $y_i=1$. Because $TP + FN$ corresponds to the number of positive samples in the batch and is independent of $p_i$, we have:
\begin{equation}
    \frac{\partial Se}{\partial p_i}
    =
    \frac{1}{TP + FN + \epsilon}
\end{equation}
Substituting this relationship into the gradient of $\mathcal{J}$ shows that the contribution of positive samples is scaled by $\frac{1}{(Se+\epsilon)(TP+FN+\epsilon)}$. When sensitivity is low, this term increases the optimization pressure on positive cases; symmetrically, when specificity is low, the corresponding negative-class gradients are amplified. In this way, the regularizer provides a soft training-time mechanism for reducing sensitivity--specificity imbalance, rather than relying solely on post-hoc threshold selection.

Our final training objective couples the CE loss, $\mathcal{L}_{\mathrm{CE}}$, with the proposed clinically aligned trade-off regularization term:
\begin{equation}
    \mathcal{L}_{\mathrm{total}}
    =
    \mathcal{L}_{\mathrm{CE}}
    +
    \lambda_{\mathrm{gm}}
    \mathcal{L}_{\mathrm{gm}},
\end{equation}
where
\begin{equation}
    \mathcal{L}_{\mathrm{gm}}
    =
    1 -
    \exp\left(
    \frac{\ln(Se+\epsilon) + \ln(Sp+\epsilon)}{2}
    \right).
\end{equation}
Since $\mathcal{L}_{\mathrm{gm}}=1-\exp(\mathcal{J}/2)$, its gradient satisfies $\nabla_\theta\mathcal{L}_{\mathrm{gm}}=-\tfrac{1}{2}\exp(\mathcal{J}/2)\nabla_\theta\mathcal{J}$. Thus, gradient descent on $\mathcal{L}_{\mathrm{gm}}$ follows the ascent direction of $\mathcal{J}$, preserving the relative class-wise weighting while rescaling the overall gradient magnitude. This scale affects its contribution when combined with CE; $\lambda_{\mathrm{gm}}$ controls the regularization strength.

\section{Experiments}
\subsection{Experimental Settings}

\begin{table}[h]
\centering
\caption{Patient-level dataset statistics. Bag size is reported as mean/median [min, max].}
\label{tab:dataset_statistics}
\footnotesize
\setlength{\tabcolsep}{5pt}
\renewcommand{\arraystretch}{1.12}
\begin{tabular}{llrrrrc}
\toprule
Dataset & Split & Images & Patients & Benign & Malignant & Bag size \\
\midrule
\multirow{2}{*}{ThyroidXL}
& Train & 9,541 & 3,354 & 2,477 & 877 & 2.85/3 [1, 10] \\
& Val   & 2,094 & 739   & 386   & 353 & 2.83/3 [1, 8]  \\
\midrule
\multirow{2}{*}{BUS-BRA}
& Train & 1,323 & 744 & 505 & 239 & 1.78/2 [1, 2] \\
& Test  & 552   & 320 & 217 & 103 & 1.73/2 [1, 2] \\
\bottomrule
\end{tabular}
\end{table}

\textbf{Datasets.} We evaluate C$^2$LUE on two biopsy-proven ultrasound benchmarks: ThyroidXL~\cite{duong2025thyroidxl}, the largest public thyroid dataset comprising 11,635 images from 4,093 patients, and BUS-BRA~\cite{gomez2024bus}, containing 1,875 breast images from 1,064 patients with multi-view scanning planes. All evaluations strictly follow the official data splits for fair and reproducible comparison. Table~\ref{tab:dataset_statistics} summarizes the patient-level statistics of both benchmarks. In our setting, each patient is represented as a bag of ultrasound images, whose size reflects the available view-level evidence for diagnosis. This formulation follows routine clinical practice, where physicians integrate multiple views for patient-level decisions, and enables us to evaluate whether a model can aggregate complementary diagnostic cues under varying bag sizes.

\noindent\textbf{Evaluation Metrics.} To align with clinical workflows, performance is evaluated strictly at the patient level. For statistical robustness, we report the mean and standard deviation of the Area Under the ROC Curve (AUC), Area Under the PR Curve (AUPRC), Sensitivity, and Specificity across five random seeds. AUC and AUPRC are reported as threshold-independent discrimination metrics. For threshold-dependent metrics, the decision threshold is selected on an internal validation subset derived from the training split by maximizing $\min(Se_{\mathrm{val}}, Sp_{\mathrm{val}})$, and is then fixed for evaluation on the official split. The same threshold-selection criterion is applied to all compared methods.

\noindent\textbf{Baselines.}
To ensure a fair and consistent comparison, all competing methods are evaluated under the same patient-level protocol. We compare C$^2$LUE with standard probability aggregation methods, including MeanP and MaxP~\cite{carbonneau2018multiple}, voting-based aggregation WMV~\cite{xiong2021comparative}, representative attention-based MIL models, including ABMIL, GatedMIL, and DSMIL~\cite{ilse2018attention,li2021dual}, and Transformer-based set aggregators, including SetTransformer and TransMIL~\cite{lee2019set,shao2021transmil}.

\noindent\textbf{Implementation Details.}
We employ a pre-trained DINOv3 ViT-B/16 backbone adapted via LoRA ($r=8$) on the attention and projection layers. All MIL baselines follow the same encoder, LoRA setting, preprocessing, optimization schedule, and patient-level batch construction, differing only in the aggregation module and classification head. Inputs are resized to $224 \times 224$ with random flip and affine augmentations. The network is trained for 100 epochs using AdamW (weight decay 0.05) and a cosine annealing learning rate ($2 \times 10^{-5}$ for the backbone, $2 \times 10^{-4}$ for task heads). Patient-level batch sizes are 32 for ThyroidXL and 64 for BUS-BRA.
\input{main_patient}

\subsection{Main Results}

\textbf{Comparison with mainstream baselines.}
As shown in Table~\ref{tab:main_patient}, C$^2$LUE achieves the strongest patient-level performance on both benchmarks, obtaining the best AUC/AUPRC of 0.932/0.919 on ThyroidXL and 0.936/0.880 on BUS-BRA. These results indicate that the proposed framework improves patient-level discrimination while remaining effective under imbalanced benign--malignant distributions.

Beyond discrimination, C$^2$LUE yields a more clinically favorable sensitivity--specificity profile. Several baselines achieve high specificity at the cost of substantially reduced sensitivity, suggesting a bias toward benign predictions or overly conservative malignancy detection. For example, SetTransformer reaches high specificity on ThyroidXL but drops to 0.507 sensitivity, indicating a considerable risk of missed malignancies. In contrast, C$^2$LUE improves sensitivity to 0.874 on ThyroidXL and 0.876 on BUS-BRA, with only a moderate specificity trade-off. This error redistribution is more consistent with screening-oriented ultrasound diagnosis, where reducing false negatives is particularly important while maintaining an acceptable false-positive rate.

\begin{figure}[htbp]
\centering
\includegraphics[width=1.0\linewidth]{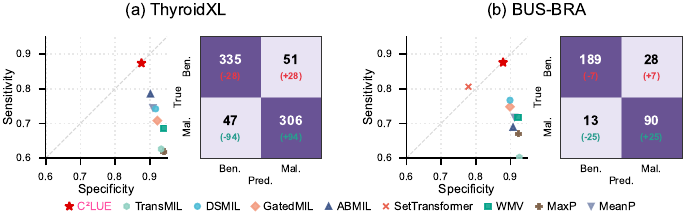}
\caption{\textbf{Decision Boundary Alignment.} Left: sensitivity--specificity trade-off. Right: representative single-run confusion matrices; parentheses denote changes relative to the baseline.}
\label{fig:fig2}
\end{figure}

\textbf{Decision boundary alignment.} Figure~\ref{fig:fig2} further visualizes the sensitivity--specificity trade-off. Compared with baselines that concentrate in specificity-dominant regions, C$^2$LUE moves the operating point closer to the balanced diagonal where sensitivity and specificity are comparable. This suggests that the proposed trade-off regularization does not merely alter threshold-level performance, but encourages the learned score distribution toward a more clinically balanced decision region during training.

The confusion matrices provide an intuitive patient-level interpretation of this shift. In the representative evaluation results, C$^2$LUE markedly reduces false negatives, indicating fewer missed malignant cases. Although this comes with an increase in false positives, the resulting error profile better reflects the clinical priority of avoiding delayed diagnosis in malignant cases. Therefore, the proposed regularization acts as a training-time mechanism for shaping the sensitivity--specificity trade-off, rather than relying entirely on post-hoc threshold adjustment.

\begin{figure}[htbp]
    \centering
    \includegraphics[width=1.0\linewidth]{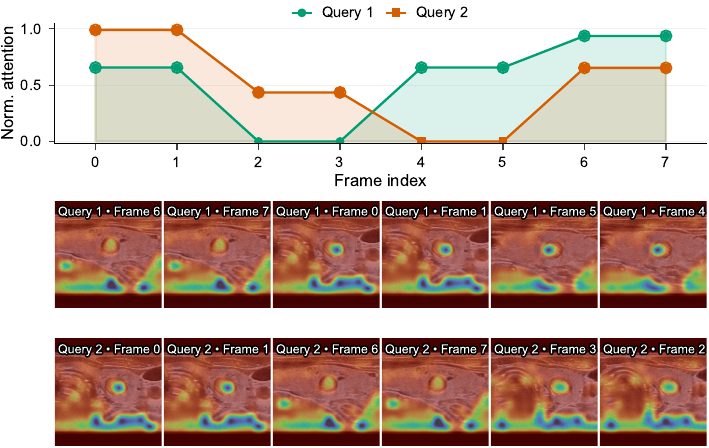}
    \caption{Multi-query evidence visualization. Top: query-specific attention of representative queries across instances. Bottom: complementary diagnostic regions highlighted by different queries.}
    \label{fig:fig3}
\end{figure}
\textbf{Evidence-aware analysis.} Figure~\ref{fig:fig3} illustrates the behavior of the multi-query aggregation module. Representative latent queries attend to different subsets of the patient bag and highlight complementary image regions. This supports the motivation that a single attention bottleneck may be insufficient for ultrasound diagnosis, where diagnostic cues such as lesion boundary irregularity, echogenicity changes, and focal suspicious patterns may appear across different views. By preserving query-specific evidence before classification, C$^2$LUE reduces evidence homogenization and provides a richer patient-level representation.

\input{abl_patient_updated}

\textbf{Ablation study.}
Table~\ref{tab:abl_patient_updated} validates the contribution of each component. For evidence aggregation, both the single-query setting and the over-parameterized query setting perform worse than the four-query design, indicating that an appropriate number of latent queries is important for balancing evidence diversity and model stability. Replacing flattened fusion with mean fusion also consistently degrades performance, suggesting that averaging query outputs weakens query-specific evidence and reintroduces feature homogenization.

The trade-off regularization ablation further confirms the role of the clinical objective. Without the trade-off regularization term ($\lambda_{\mathrm{gm}}=0$), the model tends to favor specificity at the cost of sensitivity, reflecting the imbalance induced by standard cross-entropy optimization. A moderate regularization weight improves sensitivity and yields a more balanced sensitivity--specificity profile, whereas an overly large weight can over-constrain optimization. The backbone ablations show that DINOv3 provides stronger representations than ResNet-50, ViT-B/16~\cite{dosovitskiy2021an}, and DINOv2~\cite{oquab2024dinov2}. LoRA achieves a favorable balance between patient-level performance and parameter-efficient adaptation, and is therefore adopted as the default strategy.

\begin{figure}[htbp]
    \centering
    \includegraphics[width=1.0\linewidth]{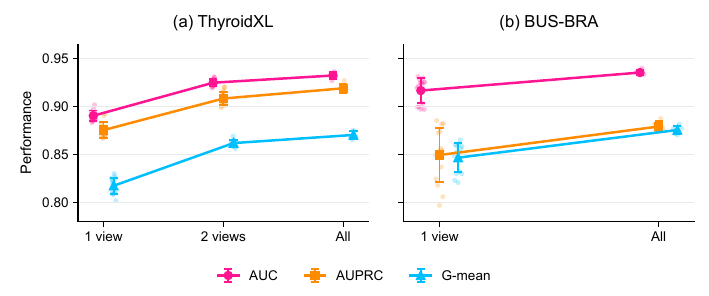}
    \caption{View-availability analysis on (a) ThyroidXL and (b) BUS-BRA. Patient-level performance improves as more ultrasound views are available. Error bars denote standard deviations over five random seeds.}
    \label{fig:fig4}
\end{figure}

\textbf{View availability analysis.}
Figure~\ref{fig:fig4} evaluates how the number of available ultrasound views affects patient-level prediction. Performance generally improves when more views are available, especially in terms of AUC and G-mean ($\sqrt{Se \cdot Sp}$). This indicates that additional views provide complementary diagnostic evidence rather than redundant information only. The trend also supports the patient-level MIL formulation: reliable ultrasound diagnosis benefits from aggregating multi-view evidence, and C$^2$LUE is able to exploit such evidence effectively under varying bag sizes.

\section{Conclusion}

We propose C$^2$LUE to address evidence homogenization and metric imbalance in patient-level ultrasound diagnosis. By coupling a LoRA-adapted DINOv3 backbone with multi-query cross-attention, C$^2$LUE disentangles heterogeneous diagnostic cues and alleviates unimodal attention collapse. In parallel, the proposed dynamic trade-off regularization objective penalizes sensitivity--specificity asymmetry during training, encouraging a more clinically balanced decision boundary. Evaluations on ThyroidXL and BUS-BRA demonstrate the best mean diagnostic performance among compared baselines together with a more balanced error profile. These results suggest that C$^2$LUE provides a promising framework for reliable patient-level ultrasound screening and diagnosis.

\begin{credits}
\subsubsection{\ackname}
This work was supported in part by the National Natural Science Foundation of China under Grant No. 62373352 and in part by the Beijing Natural Science Foundation under Grant No. M22008.
\subsubsection{\discintname}
The authors have no competing interests to declare that are relevant to the content of this article.
\end{credits}


\bibliographystyle{splncs04}

\end{document}

%% file: main_patient.tex
\begin{table}[htbp]
\centering
\caption{Main results (mean$\pm$std, 5 seeds). \DeltaMark~shows the \textcolor{green!60!black}{gain} of our method \textit{vs}. the \SND{second-best} method. \textcolor{gray}{Gray} value shows the decrease compared to the \BEST{best} method.}
\label{tab:main_patient}

\resizebox{\linewidth}{!}{
    \setlength{\tabcolsep}{1.5pt} 
    \renewcommand{\arraystretch}{1.5} 
    \begin{tabular}{lcccc|cccc}
    \toprule
    \multirow{2}{*}{\large Method} & \multicolumn{4}{c}{\large ThyroidXL} & \multicolumn{4}{c}{\large BUS-BRA} \\
    \cmidrule(lr){2-5} \cmidrule(lr){6-9}
     & \large AUC & \large AUPRC & \large $Se$ & \large $Sp$ & \large AUC & \large AUPRC & \large $Se$ & \large $Sp$ \\
    \midrule
    
    MeanP & \PM{0.912}{0.004} & \PM{0.905}{0.006} & \PM{0.745}{0.005} & \PM{0.909}{0.007} & \SND{\PM{0.929}{0.005}} & \PM{0.868}{0.004} & \PM{0.718}{0.006} & \PM{0.912}{0.008} \\
    MaxP & \PM{0.916}{0.009} & \PM{0.876}{0.007} & \PM{0.618}{0.005} & \SND{\PM{0.941}{0.004}} & \PM{0.927}{0.006} & \PM{0.865}{0.004} & \PM{0.670}{0.005} & \SND{\PM{0.923}{0.007}} \\
    WMV & \PM{0.918}{0.005} & \PM{0.874}{0.006} & \PM{0.686}{0.004} & \PM{0.940}{0.003} & \PM{0.925}{0.007} & \PM{0.869}{0.005} & \PM{0.718}{0.006} & \PM{0.922}{0.004} \\
    ABMIL & \PM{0.922}{0.006} & \SND{\PM{0.913}{0.004}} & \SND{\PM{0.785}{0.008}} & \PM{0.902}{0.005} & \PM{0.922}{0.004} & \PM{0.871}{0.007} & \PM{0.689}{0.003} & \PM{0.908}{0.006} \\
    GatedMIL & \PM{0.924}{0.004} & \PM{0.911}{0.005} & \PM{0.708}{0.007} & \PM{0.922}{0.006} & \PM{0.917}{0.005} & \PM{0.870}{0.006} & \PM{0.748}{0.004} & \PM{0.899}{0.008} \\
    DSMIL & \SND{\PM{0.927}{0.007}} & \PM{0.911}{0.008} & \PM{0.742}{0.006} & \PM{0.917}{0.005} & \SND{\PM{0.929}{0.006}} & \SND{\PM{0.875}{0.005}} & \PM{0.767}{0.007} & \PM{0.899}{0.004} \\
    SetTransformer & \PM{0.902}{0.005} & \PM{0.890}{0.006} & \PM{0.507}{0.004} & \BEST{\PM{0.946}{0.003}} & \PM{0.875}{0.006} & \PM{0.794}{0.005} & \SND{\PM{0.806}{0.004}} & \PM{0.779}{0.008} \\
    TransMIL & \PM{0.908}{0.004} & \PM{0.891}{0.007} & \PM{0.626}{0.006} & \PM{0.933}{0.005} & \PM{0.881}{0.005} & \PM{0.801}{0.008} & \PM{0.602}{0.007} & \BEST{\PM{0.926}{0.006}} \\
    C$^2$LUE (ours) & \BEST{\PM{0.932}{0.003}} & \BEST{\PM{0.919}{0.004}} & \BEST{\PM{0.874}{0.004}} & \PM{0.876}{0.003} & \BEST{\PM{0.936}{0.002}} & \BEST{\PM{0.880}{0.005}} & \BEST{\PM{0.876}{0.005}} & \PM{0.879}{0.003} \\
     & \GAIN{\textcolor{green!60!black}{$\Delta$+0.005}} & \GAIN{\textcolor{green!60!black}{+0.006}} & \GAIN{\textcolor{green!60!black}{+0.089}} & \GAIN{\textcolor{gray}{-0.070}} & \GAIN{\textcolor{green!60!black}{$\Delta$+0.007}} & \GAIN{\textcolor{green!60!black}{+0.005}} & \GAIN{\textcolor{green!60!black}{+0.070}} & \GAIN{\textcolor{gray}{-0.047}} \\
    \bottomrule
    \end{tabular}
}
\end{table}

%% file: abl_patient_updated.tex
\begin{table}[htbp]
\centering
\caption{Ablation study on patient-level performance (mean$\pm$std, 5 seeds). \DeltaMark~shows the \textcolor{green!60!black}{gain} of ours relative to the \SND{second-best} setting within each ablation block. \textcolor{gray}{Gray} value shows the decrease compared to the \BEST{best} setting.}
\label{tab:abl_patient_updated}

\resizebox{\linewidth}{!}{%
    \setlength{\tabcolsep}{1.5pt} 
    \renewcommand{\arraystretch}{1.5} 
    \begin{tabular}{lcccc|cccc}
    \toprule
    \multirow{2}{*}{\large \textbf{Setting}} & \multicolumn{4}{c}{\large ThyroidXL} & \multicolumn{4}{c}{\large BUS-BRA} \\
    \cmidrule(lr){2-5} \cmidrule(lr){6-9}
     & \large AUC & \large AUPRC & \large $Se$ & \large $Sp$ & \large AUC & \large AUPRC & \large $Se$ & \large $Sp$ \\
    
    \midrule
    \rowcolor{gray!10} \multicolumn{9}{l}{\textbf{Evidence Aggregation Strategy}} \\
    Q=1 & \SND{\PM{0.916}{0.005}} & \SND{\PM{0.902}{0.004}} & \PM{0.853}{0.006} & \PM{0.852}{0.007} & \PM{0.887}{0.004} & \PM{0.808}{0.005} & \SND{\PM{0.825}{0.006}} & \PM{0.820}{0.004} \\
    Q=8 & \PM{0.914}{0.006} & \PM{0.889}{0.005} & \SND{\PM{0.853}{0.004}} & \SND{\PM{0.855}{0.003}} & \SND{\PM{0.888}{0.007}} & \SND{\PM{0.821}{0.006}} & \PM{0.824}{0.005} & \SND{\PM{0.834}{0.008}} \\
    Q=4~\greencheck  & \BEST{\PM{0.932}{0.003}} & \BEST{\PM{0.919}{0.004}} & \BEST{\PM{0.874}{0.004}} & \BEST{\PM{0.876}{0.003}} & \BEST{\PM{0.936}{0.002}} & \BEST{\PM{0.880}{0.005}} & \BEST{\PM{0.876}{0.005}} & \BEST{\PM{0.879}{0.003}} \\
     & \GAIN{$\Delta$+0.016} & \GAIN{+0.017} & \GAIN{+0.021} & \GAIN{+0.021} & \GAIN{$\Delta$+0.048} & \GAIN{+0.059} & \GAIN{+0.051} & \GAIN{+0.045} \\
    
    \midrule
    \rowcolor{gray!10} \multicolumn{9}{l}{\textbf{Clinically-Aligned Regularization}} \\
    $\lambda_{gm}=0$ & \SND{\PM{0.912}{0.005}} & \SND{\PM{0.895}{0.007}} & \PM{0.813}{0.008} & \BEST{\PM{0.907}{0.004}} & \PM{0.887}{0.006} & \PM{0.801}{0.005} & \PM{0.816}{0.004} & \BEST{\PM{0.916}{0.006}} \\
    $\lambda_{gm}=0.4$ & \PM{0.909}{0.006} & \PM{0.891}{0.005} & \SND{\PM{0.839}{0.004}} & \PM{0.839}{0.005} & \SND{\PM{0.888}{0.004}} & \SND{\PM{0.808}{0.007}} & \SND{\PM{0.835}{0.006}} & \PM{0.825}{0.005} \\
    $\lambda_{gm}=0.2$~\greencheck  & \BEST{\PM{0.932}{0.003}} & \BEST{\PM{0.919}{0.004}} & \BEST{\PM{0.874}{0.004}} & \SND{\PM{0.876}{0.003}} & \BEST{\PM{0.936}{0.002}} & \BEST{\PM{0.880}{0.005}} & \BEST{\PM{0.876}{0.005}} & \SND{\PM{0.879}{0.003}} \\
     & \GAIN{$\Delta$+0.020} & \GAIN{+0.024} & \GAIN{+0.035} & {\color{gray}-0.031} & \GAIN{$\Delta$+0.048} & \GAIN{+0.072} & \GAIN{+0.041} & {\color{gray}-0.037} \\
    
    \midrule
    \rowcolor{gray!10} \multicolumn{9}{l}{\textbf{Feature Fusion}} \\
    meanQ & \SND{\PM{0.912}{0.004}} & \SND{\PM{0.899}{0.006}} & \SND{\PM{0.850}{0.007}} & \SND{\PM{0.850}{0.005}} & \SND{\PM{0.891}{0.005}} & \SND{\PM{0.813}{0.004}} & \SND{\PM{0.825}{0.006}} & \SND{\PM{0.830}{0.007}} \\
    flat~\greencheck  & \BEST{\PM{0.932}{0.003}} & \BEST{\PM{0.919}{0.004}} & \BEST{\PM{0.874}{0.004}} & \BEST{\PM{0.876}{0.003}} & \BEST{\PM{0.936}{0.002}} & \BEST{\PM{0.880}{0.005}} & \BEST{\PM{0.876}{0.005}} & \BEST{\PM{0.879}{0.003}} \\
     & \GAIN{$\Delta$+0.020} & \GAIN{+0.020} & \GAIN{+0.024} & \GAIN{+0.026} & \GAIN{$\Delta$+0.045} & \GAIN{+0.067} & \GAIN{+0.051} & \GAIN{+0.049} \\
    
    \midrule
    \rowcolor{gray!10} \multicolumn{9}{l}{\textbf{Backbone Adaptation}} \\
    frz & \PM{0.877}{0.004} & \PM{0.858}{0.006} & \PM{0.799}{0.005} & \PM{0.801}{0.008} & \PM{0.882}{0.007} & \PM{0.799}{0.004} & \PM{0.816}{0.006} & \PM{0.816}{0.005} \\
    full & \SND{\PM{0.926}{0.005}} & \SND{\PM{0.915}{0.004}} & \SND{\PM{0.867}{0.006}} & \SND{\PM{0.863}{0.007}} & \BEST{\PM{0.942}{0.006}} & \BEST{\PM{0.893}{0.005}} & \BEST{\PM{0.884}{0.004}} & \BEST{\PM{0.885}{0.006}} \\
    LoRA~\greencheck  & \BEST{\PM{0.932}{0.003}} & \BEST{\PM{0.919}{0.004}} & \BEST{\PM{0.874}{0.004}} & \BEST{\PM{0.876}{0.003}} & \SND{\PM{0.936}{0.002}} & \SND{\PM{0.880}{0.005}} & \SND{\PM{0.876}{0.005}} & \SND{\PM{0.879}{0.003}} \\
     & \GAIN{$\Delta$+0.006} & \GAIN{+0.004} & \GAIN{+0.007} & \GAIN{+0.013} & {\color{gray}-0.006} & {\color{gray}-0.013} & {\color{gray}-0.008} & {\color{gray}-0.006} \\
    
    \midrule
    \rowcolor{gray!10} \multicolumn{9}{l}{\textbf{Backbone Architecture}} \\ 
    ResNet-50 & \PM{0.921}{0.004} & \PM{0.904}{0.006} & \PM{0.866}{0.010} & \PM{0.855}{0.009} & \PM{0.918}{0.003} & \PM{0.852}{0.005} & \PM{0.867}{0.005} & \PM{0.862}{0.004} \\
    ViT-B/16 & \PM{0.924}{0.004} & \PM{0.911}{0.006} & \PM{0.862}{0.010} & \PM{0.860}{0.009} & \PM{0.922}{0.003} & \PM{0.867}{0.005} & \PM{0.859}{0.005} & \PM{0.861}{0.004} \\
    DINOv2  & \SND{\PM{0.928}{0.004}} & \SND{\PM{0.915}{0.006}} & \SND{\PM{0.869}{0.010}} & \SND{\PM{0.870}{0.009}} & \SND{\PM{0.932}{0.003}} & \SND{\PM{0.876}{0.005}} & \SND{\PM{0.868}{0.005}} & \SND{\PM{0.871}{0.004}} \\
    DINOv3~\greencheck & \BEST{\PM{0.932}{0.003}} & \BEST{\PM{0.919}{0.004}} & \BEST{\PM{0.874}{0.004}} & \BEST{\PM{0.876}{0.003}} & \BEST{\PM{0.936}{0.002}} & \BEST{\PM{0.880}{0.005}} & \BEST{\PM{0.876}{0.005}} & \BEST{\PM{0.879}{0.003}} \\
     & \GAIN{$\Delta$+0.004} & \GAIN{+0.004} & \GAIN{+0.005} & \GAIN{+0.006} & \GAIN{$\Delta$+0.004} & \GAIN{+0.004} & \GAIN{+0.008} & \GAIN{+0.008} \\
    \bottomrule
    \end{tabular}
}
\end{table}

%% file: samplepaper.bbl
\begin{thebibliography}{99}

\bibitem{van2019deep}
Van Sloun, R.J.G., Cohen, R., Eldar, Y.C.: Deep learning in ultrasound imaging. Proceedings of the IEEE \textbf{108}(1), 11--29 (2020)

\bibitem{qian2021prospective}
Qian, X., Pei, J., Zheng, H., Xie, X., Yan, L., Zhang, H., Han, C., Gao, X., Zhang, H., Zheng, W., et al.: Prospective assessment of breast cancer risk from multimodal multiview ultrasound images via clinically applicable deep learning. Nature Biomedical Engineering \textbf{5}(6), 522--532 (2021)

\bibitem{AhmadiAmiri2024TMI_AS}
Ahmadi Amiri, S.N., Tsang, M.Y., Gu, A.N., Tsang, T.S.M., Abolmaesumi, P.:
Transformer-Based Spatio-Temporal Analysis for Classification of Aortic Stenosis Severity From Echocardiography Cine Series.
IEEE Transactions on Medical Imaging \textbf{43}(1), 366--376 (2024)

\bibitem{wan2021hierarchical}
Wan, P., Chen, F., Liu, C., Kong, W., Zhang, D.: Hierarchical temporal attention network for thyroid nodule recognition using dynamic CEUS imaging. IEEE Transactions on Medical Imaging \textbf{40}(6), 1646--1660 (2021)

\bibitem{freitas2024automatic}
Freitas, J., Gomes-Fonseca, J., Tonelli, A.C., Correia-Pinto, J., Fonseca, J.C., Queir{\'o}s, S.: Automatic multi-view pose estimation in focused cardiac ultrasound. Medical Image Analysis \textbf{94}, 103146 (2024)

\bibitem{tao2025enhancing}
Tao, X., Cao, Y., Jiang, Y., Wu, X., Yan, D., Xue, W., Zhuang, S., Yang, X., Huang, R., Zhang, J., et al.: Enhancing lesion detection in automated breast ultrasound using unsupervised multi-view contrastive learning with 3D DETR. Medical Image Analysis \textbf{101}, 103466 (2025)

\bibitem{maron1997framework}
Maron, O., Lozano-P{\'e}rez, T.: A framework for multiple-instance learning. Advances in Neural Information Processing Systems \textbf{10} (1997)

\bibitem{carbonneau2018multiple}
Carbonneau, M.-A., Cheplygina, V., Granger, E., Gagnon, G.: Multiple instance learning: A survey of problem characteristics and applications. Pattern Recognition \textbf{77}, 329--353 (2018)

\bibitem{ilse2018attention}
Ilse, M., Tomczak, J., Welling, M.: Attention-based deep multiple instance learning. In: International Conference on Machine Learning, pp. 2127--2136. PMLR (2018)

\bibitem{lee2019set}
Lee, J., Lee, Y., Kim, J., Kosiorek, A., Choi, S., Teh, Y.W.: Set transformer: A framework for attention-based permutation-invariant neural networks. In: International Conference on Machine Learning, pp. 3744--3753. PMLR (2019)

\bibitem{shao2021transmil}
Shao, Z., Bian, H., Chen, Y., Wang, Y., Zhang, J., Ji, X., et al.: TransMIL: Transformer based correlated multiple instance learning for whole slide image classification. Advances in Neural Information Processing Systems \textbf{34}, 2136--2147 (2021)

\bibitem{duong2025thyroidxl}
Duong, V.H., Vu, H., Phan, H.D., Nguyen, D.Q., Pham, D.H., Le, Q.T., Nguyen, B.S., Do, T.D., Dinh, V.S., Nguyen, T.C., et al.: ThyroidXL: Advancing thyroid nodule diagnosis with an expert-labeled, pathology-validated dataset. In: Medical Image Computing and Computer Assisted Intervention -- MICCAI 2025, pp. 616--626. Springer (2025)

\bibitem{liang2025medical}
Liang, H., Xu, J., Zhang, Y., Huang, Y., Zhang, Y., Yang, X., Li, R., Deng, X., Liu, Y., Tao, G., et al.: Medical-knowledge driven multiple instance learning for classifying severe abdominal anomalies on prenatal ultrasound. In: Medical Image Computing and Computer Assisted Intervention -- MICCAI 2025, Lecture Notes in Computer Science, vol. 15962, pp. 344--354. Springer Nature Switzerland (2025)

\bibitem{Wang2020MedIA_Thyroid}
Wang, L., Zhang, L., Zhu, M., Qi, X., Yi, Z.:
Automatic diagnosis for thyroid nodules in ultrasound images by deep neural networks.
Medical Image Analysis \textbf{61}, 101665 (2020)

\bibitem{Zhao2022TMI_LGFD}
Zhao, S.X., Chen, Y., Yang, K.F., Luo, Y., Ma, B.Y., Li, Y.J.:
A Local and Global Feature Disentangled Network: Toward Classification of Benign-Malignant Thyroid Nodules From Ultrasound Image.
IEEE Transactions on Medical Imaging \textbf{41}(6), 1497--1509 (2022)

\bibitem{ren2020balanced}
Ren, J., Yu, C., Ma, X., Zhao, H., Yi, S., et al.: Balanced meta-softmax for long-tailed visual recognition. Advances in Neural Information Processing Systems \textbf{33}, 4175--4186 (2020)

\bibitem{yuan2021large}
Yuan, Z., Yan, Y., Sonka, M., Yang, T.: Large-scale robust deep AUC maximization: A new surrogate loss and empirical studies on medical image classification. In: Proceedings of the IEEE/CVF International Conference on Computer Vision, pp. 3040--3049 (2021)

\bibitem{nelson2016harms}
Nelson, H.D., Pappas, M., Cantor, A., Griffin, J., Daeges, M., Humphrey, L.: Harms of breast cancer screening: Systematic review to update the 2009 US Preventive Services Task Force recommendation. Annals of Internal Medicine \textbf{164}(4), 256--267 (2016)

\bibitem{menon2021longtail}
Menon, A.K., Jayasumana, S., Rawat, A.S., Jain, H., Veit, A., Kumar, S.: Long-tail learning via logit adjustment. In: International Conference on Learning Representations (2021)

\bibitem{Huang2022MICCAI_PersonalizedThyroid}
Huang, H., Dong, Y., Jia, X., Zhou, J., Ni, D., Cheng, J., Huang, R.:
Personalized diagnostic tool for thyroid cancer classification using multi-view ultrasound.
In: Medical Image Computing and Computer Assisted Intervention -- MICCAI 2022, Lecture Notes in Computer Science, pp. 665--674. Springer (2022)

\bibitem{Yan2024CommMed_BreastUS}
Yan, L., Liang, Z., Zhang, H., Zhang, G., Zheng, W., Han, C., Yu, D., Zhang, H., Xie, X., Liu, C., et al.: A domain knowledge-based interpretable deep learning system for improving clinical breast ultrasound diagnosis. Communications Medicine \textbf{4}, 90 (2024)

\bibitem{Lu2021NBE_CLAM}
Lu, M.Y., Williamson, D.F.K., Chen, T.Y., Chen, R.J., Barbieri, M., Mahmood, F.: Data-efficient and weakly supervised computational pathology on whole-slide images. Nature Biomedical Engineering \textbf{5}(6), 555--570 (2021)

\bibitem{li2021dual}
Li, B., Li, Y., Eliceiri, K.W.: Dual-stream multiple instance learning network for whole slide image classification with self-supervised contrastive learning. In: Proceedings of the IEEE/CVF Conference on Computer Vision and Pattern Recognition, pp. 14318--14328 (2021)

\bibitem{Guo2017ICML_Calibration}
Guo, C., Pleiss, G., Sun, Y., Weinberger, K.Q.: On calibration of modern neural networks. In: International Conference on Machine Learning, pp. 1321--1330. PMLR (2017)

\bibitem{Rajaraman2022PLOSONE_Calibration}
Rajaraman, S., Ganesan, P., Antani, S.: Deep learning model calibration for improving performance in class-imbalanced medical image classification tasks. PLOS ONE \textbf{17}(1), e0262838 (2022)

\bibitem{Geysels2025MICCAI_CalibrationMIL}
Geysels, A., Van Calster, B., De Moor, B., Froyman, W., Timmerman, D.: Calibration in multiple instance learning: Evaluating aggregation methods for ultrasound-based diagnosis. In: Medical Image Computing and Computer Assisted Intervention -- MICCAI 2025, Lecture Notes in Computer Science, vol. 15974, pp. 55--64. Springer Nature Switzerland (2025)

\bibitem{Scholz2024MIDL_ImbalanceLoss}
Scholz, D., Erdur, A.C., Buchner, J.A., Peeken, J.C., Rueckert, D., Wiestler, B.: Imbalance-aware loss functions improve medical image classification. In: Medical Imaging with Deep Learning, Proceedings of Machine Learning Research, vol. 250, pp. 1341--1356. PMLR (2024)



\bibitem{simeoni2025dinov3}
Sim{\'e}oni, O., Vo, H.V., Seitzer, M., Baldassarre, F., Oquab, M., Jose, C., Khalidov, V., Szafraniec, M., Yi, S., Ramamonjisoa, M., et al.: DINOv3. arXiv preprint arXiv:2508.10104 (2025)

\bibitem{hu2022lora}
Hu, E.J., Shen, Y., Wallis, P., Allen-Zhu, Z., Li, Y., Wang, S., Wang, L., Chen, W., et al.: LoRA: Low-rank adaptation of large language models. In: International Conference on Learning Representations (2022)

\bibitem{gomez2024bus}
G{\'o}mez-Flores, W., Gregorio-Calas, M.J., Coelho de Albuquerque Pereira, W.: BUS-BRA: A breast ultrasound dataset for assessing computer-aided diagnosis systems. Medical Physics \textbf{51}(4), 3110--3123 (2024)

\bibitem{xiong2021comparative}
Xiong, D., Zhang, Z., Wang, T., Wang, X.: A comparative study of multiple instance learning methods for cancer detection using T-cell receptor sequences. Computational and Structural Biotechnology Journal \textbf{19}, 3255--3268 (2021)

\bibitem{dosovitskiy2021an}
Dosovitskiy, A., Beyer, L., Kolesnikov, A., Weissenborn, D., Zhai, X., Unterthiner, T.,
Dehghani, M., Minderer, M., Heigold, G., Gelly, S., Uszkoreit, J., Houlsby, N.:
An image is worth 16x16 words: Transformers for image recognition at scale.
In: International Conference on Learning Representations (2021)

\bibitem{oquab2024dinov2}
Oquab, M., Darcet, T., Moutakanni, T., Vo, H.V., Szafraniec, M., Khalidov, V., Fernandez, P.,
Haziza, D., Massa, F., El-Nouby, A., Assran, M., Ballas, N., Galuba, W., Howes, R.,
Huang, P.-Y., Li, S.-W., Misra, I., Rabbat, M., Sharma, V., Synnaeve, G., Xu, H.,
Jegou, H., Mairal, J., Labatut, P., Joulin, A., Bojanowski, P.:
DINOv2: Learning robust visual features without supervision.
Transactions on Machine Learning Research (2024)



\end{thebibliography}
